\documentclass[prl,twocolumn,showpacs,amsmath,amssymb]{revtex4}

\usepackage{bm}
\usepackage[dvipdfm]{graphicx}
\usepackage{color}
\renewcommand{\(}     {\left(}
\renewcommand{\)}     {\right)}
\newcommand  {\eqn}[1]{(\ref{eqn:#1})}
\def \braket<#1>{\langle{#1}\rangle}

\begin{document}

\preprint{preprint}

\title{Density of States Scaling at the Semimetal to Metal Transition in Three Dimensional Topological Insulators}
\author{Koji Kobayashi$^1$}
\author{Tomi Ohtsuki$^1$}
\author{Ken-Ichiro Imura$^{2,4}$}
\author{Igor F. Herbut$^{3,4}$}
\affiliation{$^1$Department of Physics, Sophia University, Chiyoda-ku, Tokyo, 102-8554, Japan}
\affiliation{$^2$Department of Quantum Matter, AdSM, Hiroshima University, Higashi-Hiroshima, 739-8530, Japan}
\affiliation{$^3$Department\!~of\!~Physics,\!~Simon\!~Fraser\!~University,\!~Burnaby,\!~British\!~Columbia,\!~V5A\!~1S6,\!~Canada}
\affiliation{$^4$Max-Planck-Institut f\"ur Physik komplexer Systeme, N\"othnitzer Str.~38, 01187 Dresden, Germany}

\date{\today}

\begin{abstract}
 The quantum phase transition between the three dimensional Dirac semimetal and the diffusive metal can be induced by increasing disorder.
 Taking the system of disordered $\mathbb{Z}_2$ topological insulator as an important example, 
we compute the single particle density of states by the kernel polynomial method.
 We focus on three regions:
the Dirac semimetal at the phase boundary between two topologically distinct phases, 
the tricritical point of the two topological insulator phases and the diffusive metal, 
and the diffusive metal lying at strong disorder.
 The density of states obeys a novel single parameter scaling, 
collapsing onto two branches of a universal scaling function, 
which correspond to the Dirac semimetal and the diffusive metal.
 The diverging length scale critical exponent $\nu$ and the dynamical critical exponent $z$ are estimated,
and found to differ significantly from those for the conventional Anderson transition.
 Critical behavior of experimentally observable quantities near and at the tricritical point is also discussed.
\end{abstract}

\pacs{
71.30.+h, 
05.70.Jk, 
71.23.-k, 
71.55.Ak  
}
\maketitle


 Topological classification of different insulating phases \cite{Schnyder08, Kitaev09} is an emerging new paradigm in condensed matter physics.
 Unlike in the Landau theory of phase transitions that is rooted in the idea of spontaneous breaking of symmetry \cite{HerbutBook}, 
it is less clear 
how to describe different universality classes of the transitions between topologically different phases.
 This is because the usual notion of the local order parameter characterizing the different phases 
is often lacking.
 At the transition between topologically distinct phases, on the other hand, the gap closes, and the system becomes a  semimetal.
 In three dimensions (3D) such a critical phase is stable in presence of weak disorder \cite{Shindou09}, 
but as disorder is increased it gives way to a diffusive metallic state \cite{Fradkin86}.
 This transition belongs to a distinct universality class that exhibits non-trivial dynamical and diverging length scale exponents $z$ and $\nu$, for example \cite{Fradkin86, GC}.
 The 3D Dirac Hamiltonian in presence of disorder is ubiquitous: 
it applies to certain phases of superfluid $^3$He \cite{Volovik03}, degenerate semiconductors \cite{Fradkin86}, 
and to the Weyl semimetals \cite{Burkov11a,Burkov11b,Neupane13,Borisenko13}.
 Related theories of disordered critical points for two-dimensional interacting Dirac fermions and bosons were also advanced in the past \cite{Igor1, Igor2}.

 In this paper we discuss how this disorder-induced fermionic criticality is reflected 
in the scaling behavior 
of a readily available physical quantity, the single particle density of states (DOS), 
which can be understood as a proper order parameter 
that characterizes such a transition.
 We then express the critical behavior of Dirac electron velocity, diffusion coefficient, conductivity and anomalous diffusion exponent in terms of $z$ and $\nu$.
 Such a surprisingly simple description is contrasted with the conventional Anderson transition \cite{anderson58,kramer93,evers08},
where the DOS remains smooth through the transition.

\begin{figure}[tb]
\includegraphics[width=85mm]{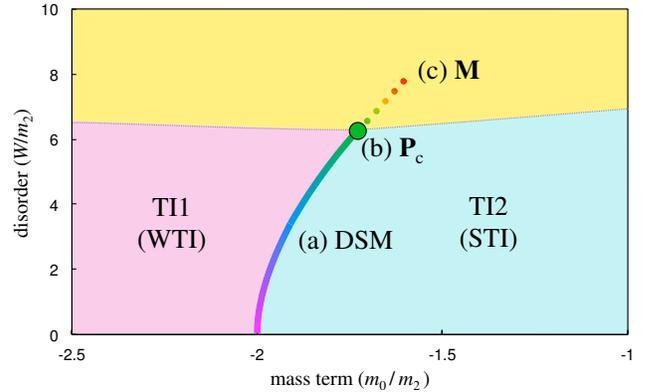}
\vspace{-2mm}
\caption{(Color online)
         Typical phase diagram of the system under consideration.
         TI1 and TI2 correspond, respectively, to weak and strong topological insulator (WTI and STI), 
        and DSM to the critical Dirac semimetal phase.
         The dotted line in the diffusive metal (${\bf M}$) phase (c) is an extrapolation \cite{OrderOfPolynomial} of the DSM line (a).
         The tricritical point (b) is denoted as ${\bf P}_{\rm c}$.}
\label{fig:PD}
\vspace{-5mm}
\end{figure}

 In order to produce and control the semimetallic phase, 
we focus on a 3D time-reversal symmetric topological insulator under disorder.
 The $\mathbb{Z}_2$ topological insulator is interesting in itself, 
and has lately been a subject of intense theoretical and experimental research, with a number of real material realizations \cite{Hasan10}.
 Consider the phase diagram of a system exhibiting both weak and strong topological insulators (WTI and STI) 
as some parameter is varied \cite{3DTI1, 3DTI2, 3DTI3} (see Fig.~\ref{fig:PD}).
 In three spatial dimensions disorder is irrelevant in the renormalization group sense, 
so that at weak disorder a {\it direct} transition between two topologically distinct insulating phases \cite{Shindou09}, say, between TI1 and TI2, remains.
 (In the specific situation we consider below, TI1 $=$ WTI and TI2 $=$ STI.)
 Only above a finite strength of disorder $W>0$, 
does the bulk energy gap become completely filled with impurity levels,
so that the insulating phases are replaced by a diffusive metallic (${\bf M}$) phase \cite{KOI} (see Fig.~\ref{fig:PD}).
 Since TI1 and TI2 are characterized by a different topological number protected by the bulk energy gap,
at the phase boundary the bulk spectrum is in general closed.
 In the present case the system is also protected by time-reversal symmetry, 
and such a gap closing appears as a (Kramers) degenerate pair of point nodes, 
i.e., as the Dirac semimetal (DSM) \cite{Young12} line in the phase diagram.
 As disorder is increased the DSM line also terminates at the intersection with the insulator-metal phase boundary.
 In the following we focus on the evolution of the DOS as one moves along the DSM line, 
through the tricritical point ${\bf P}_{\rm c}$ where the DSM line terminates, 
and finally reaches inside the metallic phase.

\begin{figure}[tb]
\includegraphics[width=85mm]{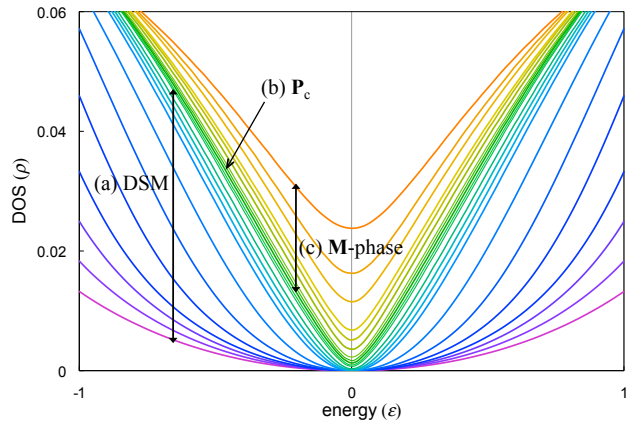}
\vspace{-2mm}
\caption{(Color online) Density of states
         calculated at different points of the phase diagram ($2\le W \le 7.5$);
         (a) on the WTI/STI boundary, 
         (b) at the tricritical point,
         and 
         (c) in the ${\bf M}$-phase.
          Its energy dependence $\rho (\epsilon)$ 
         is quadratic on the WTI/STI boundary (a), 
         becoming almost linear at the tricritical point (b),
         while it acquires a finite value $\rho (0)$ at $\epsilon =0$ on the ${\bf M}$-side (c).
         We emphasize that these DOSs are not of the surface, but of the bulk.
}
\label{fig:dos}
\end{figure}

 We have previously established, by a detailed numerical study of the conductance \cite{KOI}, 
that although disorder $W$ shifts the position of the phase boundary \cite{TAI1, TAI2, TAI3, TAI4, TAI5, TAI6} 
(determined, e.g., by the position of the conductance peak), 
it is nevertheless irrelevant;
the peak height of the conductance on the DSM line is not influenced by the disorder strength.
 It was also found \cite{KOI} that on the DSM line the DOS remains a {\it quadratic} function of low energies, exactly as in the clean limit [see the curves (a) in Fig.~\ref{fig:dos}].
 Whereas the quadratic behavior is left intact by disorder, the coefficient of the quadratic term, which is related to the velocity $v$ of Dirac electrons, is renormalized \cite{Wu12}, as in Eq.~\eqn{velocity} below.

 In this Letter we further quantify the behavior of the DOS on the DSM line toward the diffusive metal phase, 
and demonstrate that the DOS obeys a single parameter scaling typical of second order phase transitions, 
with new values of critical exponents.
 Our analysis is based on the single parameter scaling hypothesis, 
which is substantially supported by numerical results.
 The scaling behavior of the DOS is studied 
using the kernel polynomial method (KPM) \cite{Weisse06}.

 The 3D disordered $\mathbb{Z}_2$ topological insulator is modeled 
as a Wilson-Dirac-type tight-binding Hamiltonian 
with an effective momentum-dependent mass term \cite{Liu10},
\begin{align}
   m(\bm k) = m_0 + m_2 \sum_{\mu=x,y,z} (1-\cos k_\mu)\, ,
 \label{mass}
\end{align} 
implemented on a cubic lattice.
 The topological nature of the model is controlled by the ratio of two mass parameters
$m_0$ and $m_2$
such that
an STI phase with $\mathbb{Z}_2$ (one strong and three weak) indices \cite{3DTI1, 3DTI2, 3DTI3}
$(\nu_0, \nu_1 \nu_2 \nu_3)=(1, 000)$ 
appears when $-2 < m_0/m_2 < 0$,
while the regime of parameters: $-4 < m_0/m_2 < -2$
falls on a WTI phase with $(\nu_0, \nu_1 \nu_2 \nu_3)=(0, 111)$ 
(see Fig.~\ref{fig:PD}).

In real space our tight-binding Hamiltonian reads
\begin{align}
 \label{eqn:ham}
   H = & \sum_{\bm r} \sum_{\mu=x,y,z} 
   \left[
      |{\bm r}+{\bm e}_\mu \rangle
      \left(
         \frac{{\rm i}t}{2}\gamma_{\mu} -\frac{m_2}{2} \gamma_0
      \right)
      \langle {\bm r}|
      + \rm{h.c.}
   \right]  
\nonumber \\
      & +\sum_{\bm r} 
      |{\bm r} \rangle
      \Big[
         (m_0+3 m_2) \gamma_0
         + V_{\bm r} 1_4
      \Big]
      \langle {\bm r}|\, ,
\end{align}
where
${\bm e}_{\mu}$ is a unit vector in the $\mu$-direction, 
and
$1_4$ represents the $4\times4$ identity matrix.
$\gamma_{\mu}$ and $\gamma_0$ form a set of $\gamma$-matrices 
in a $4\times4$ representation,
   \begin{align} \label{eqn:gammaMat}
      \gamma_{\mu} = \begin{pmatrix}
                         0     & \sigma_\mu \\
                      \sigma_\mu &    0
                   \end{pmatrix}\, , \ 
      \gamma_0 = \begin{pmatrix}
                      1_{2} & 0 \\
                      0 & -1_{2}
                   \end{pmatrix}\, ,
   \end{align}
where $\sigma_{\mu}$ are Pauli matrices and $1_{2}$ is $2\times 2$ identity matrix.
$m_0, m_2$ and $t$ are mass and hopping parameters, 
and 
$V_{\bm r}$ represents a potential disorder
distributed uniformly and independently between $-W/2$ and $W/2$.

 For simplicity, we have assumed the Hamiltonian Eq.~\eqn{ham} to be isotropic.
 In the actual computation
we set the mass and hopping parameters to $m_2 =1$, $t=2$.
 The linear size of the system $L$ is taken to be $200$ times the lattice constant,
which is enough to reach the thermodynamic limit of DOS per unit volume.
 We also take the average over two samples, 
although the statistical error is already sufficiently small for $L=200$, 
because of the self-averaging nature of DOS.
 The order of the Chebyshev expansion in KPM is typically a few thousand, so that the DOS becomes smooth.
 The periodic boundary conditions are imposed on each direction.

 The scaling form of the density of states per volume near the Dirac point may be derived as follows.
 Begin with a dimensionless quantity, 
the number of states $N (\epsilon, L)$ below the energy $\epsilon$ in the system of size $L$ in $d$ dimensions, 
and assume that it is a function of dimensionless parameters $L/\xi$ and $\epsilon/\epsilon_0$, 
\begin{align}
   N(\epsilon, L) = F\( {L}/{\xi} , {\epsilon}/{\epsilon_0} \)\, ,
\end{align}
where $\xi$ is the characteristic length scale 
and $\epsilon_0$ is the characteristic energy scale.
 They are related via the dynamical exponent $z$,
\begin{align}
 \epsilon_0 \propto \xi^{-z}\, . 
\end{align}
 Since the number of states should be proportional to $L^d$, 
the above scaling form should be
\begin{align} \label{eqn:scalingN}
 N(\epsilon, L) = (L/\xi)^{d} f\( \epsilon \xi^{z} \)\, .
\end{align}

 From $N(\epsilon,L)$, the DOS per volume $\rho(\epsilon)$ is calculated as
\begin{align}
 \rho(\epsilon) = \frac{1}{L^d} \frac{d N(\epsilon, L)}{d \epsilon}\, ,
\end{align}
so that we finally obtain its scaling form,
\begin{align} \label{eqn:rho_xi}
 \rho(\epsilon) = \rho(-\epsilon) = \xi^{z-d} f'(|\epsilon|\xi^z)\, .
\end{align}
 The first equality comes from the symmetry of DOS about $\epsilon = 0$.
 Upon introducing the distance from the tricritical point $\delta = |W-W_{\rm c}|/W_{\rm c}$, 
we may assume that the length scale $\xi$ diverges near the tricritical point ${\bf P}_{\rm c}$ as,
\begin{align} \label{eqn:delta}
 \xi \sim \delta^{-\nu}\, ,
\end{align}
where $\nu$ is the critical exponent.
 Around ${\bf P}_{\rm c}$, 
the scaling law, Eq.~\eqn{rho_xi}, therefore reads,
\begin{align} \label{eqn:DOS_scaling_delta}
 \rho(\epsilon) \sim \delta^{(d-z)\nu} f'(|\epsilon|\delta^{-z\nu})\, .
\end{align}
 For $\epsilon \to 0$, i.e., when the argument of the scaling function is small, 
one expects qualitatively different behavior in the ${\bf M}$-phase and on the DSM line.
 If the system has Dirac cones, the DOS is expected to be proportional to $|\epsilon|^{d-1}$ for $|\epsilon| \ll \epsilon_0$, so 
\begin{align} \label{eqn:DOS_scaling_DSM}
 \rho(\epsilon) \sim \delta^{(d-z)\nu} (|\epsilon|\delta^{-z\nu})^{d-1}
 = |\epsilon|^{d-1} \delta^{-(z-1)d\nu}\, .
\end{align}
 In the ${\bf M}$-phase, on the other hand, the DOS is finite at $\epsilon=0$, and
\begin{align} \label{eqn:DOS_scaling_M}
 \rho(0) \sim \delta^{(d-z)\nu} (|\epsilon|\delta^{-z\nu})^0
 = \delta^{(d-z)\nu}\, .
\end{align}
 Right at the tricritical point $\delta=0$,
$\xi$ dependences in the prefactor and the argument of Eq.~\eqn{rho_xi} should cancel, 
and consequently, 
\begin{align}
\label{eqn:criticalPoint}
 \rho(\epsilon) \sim \delta^{(d-z)\nu} (|\epsilon|\delta^{-z\nu})^{(d-z)/z}
 = |\epsilon|^{(d-z)/z}\, .
\end{align}

\begin{figure}[tb]
  \begin{tabular}{cc}
    \includegraphics[width=40mm]{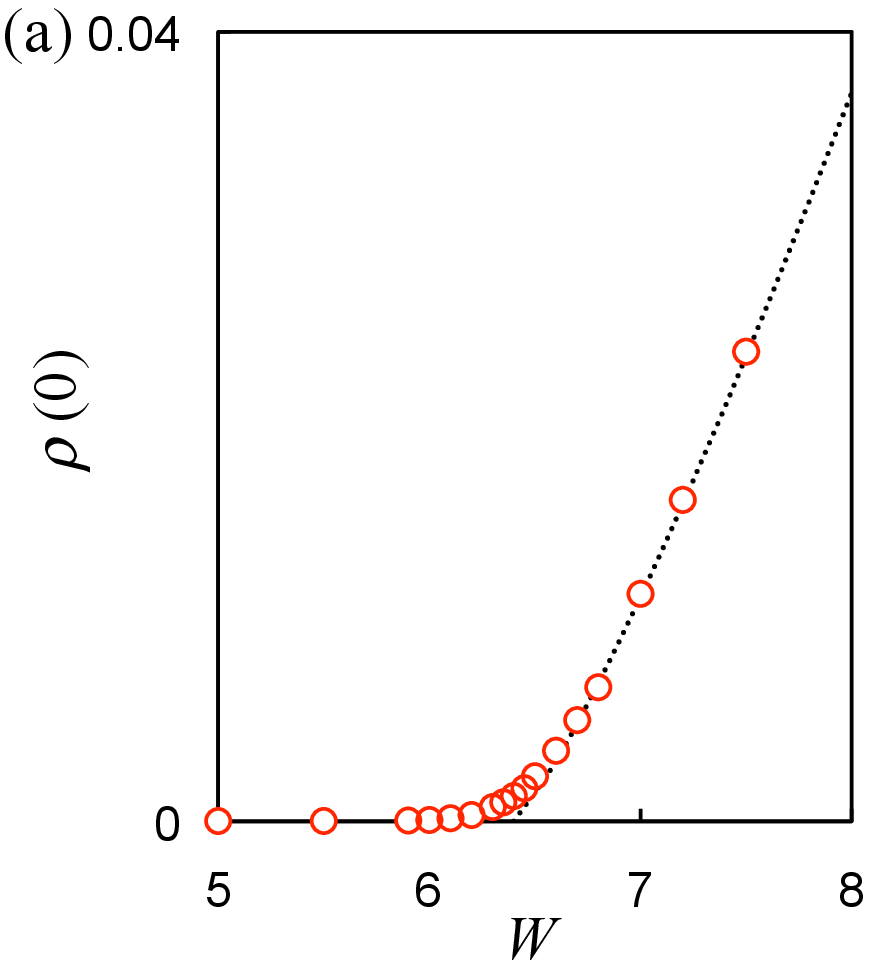} &
    \includegraphics[width=40mm]{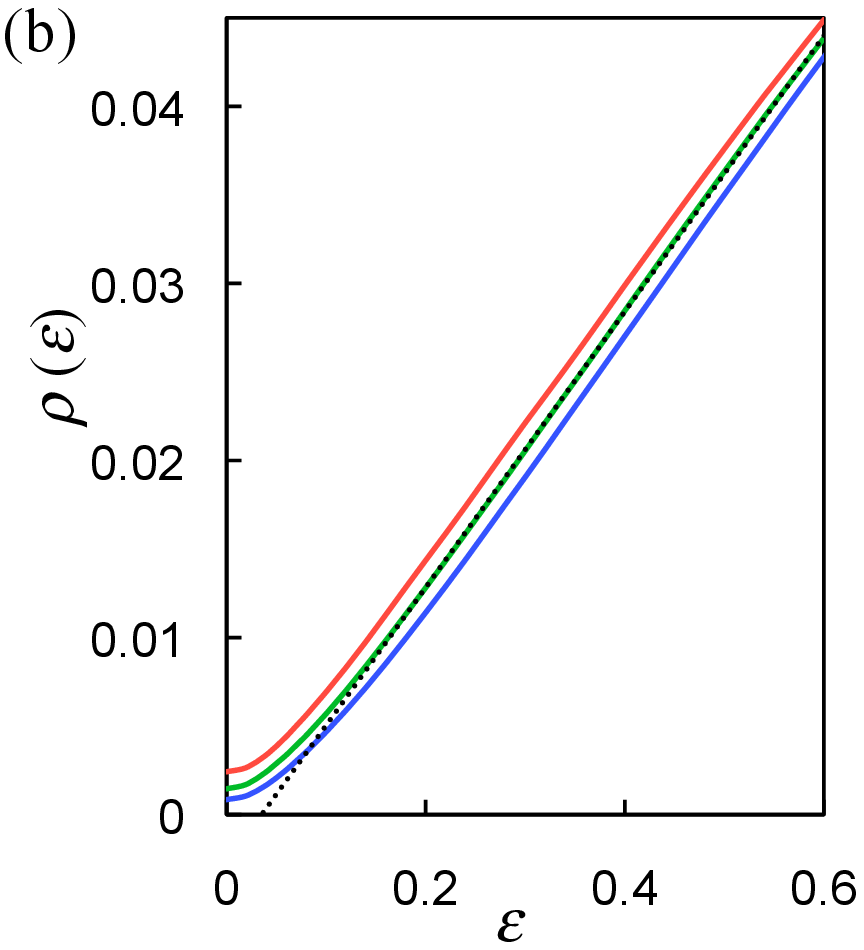}
  \end{tabular}
  \vspace{-2mm}
  \caption{ (Color online) 
     (a) The DOS at $\epsilon=0$.
     The point $W_{\rm c}$ where $\rho (0) \to 0$ indicates the tricritical point ${\bf P}_{\rm c}$. 
     (b) The DOSs around $W_{\rm c}$ (solid lines, $W=6.3,\ 6.4,\ 6.5$ from bottom to top).
     They can be approximated by a linear function (dotted line).
     The deviations for small energy regions are coming from the finite size effect $\rho(0)\sim L^{-2}$.
     We note that the effect of long ranged disorder \cite{Nandkishore13}, 
    which might survive due to the finite lattice spacing, is not identified in our numerics.
  }
  \label{fig:fit_Wc_z}
 \end{figure}

 Armed with the above observations, 
we next study the DOS numerically.
 First, the DOS at $\epsilon=0$ vanishes [Fig.~\ref{fig:fit_Wc_z}(a)] around
\begin{align} \label{eqn:Wc}
 W_{\rm c} &= 6.4 \pm 0.1\, .
\end{align}
 We use this value to define $\delta$.
 The DOSs around $W = W_{\rm c}$, i.e., near ${\bf P}_{\rm c}$,
are plotted in Fig.~\ref{fig:fit_Wc_z}(b).
 From the observed energy dependence and Eq.~\eqn{criticalPoint}, we estimate
\begin{align} \label{eqn:z}
 (3-z)/z &= 1.00 \pm 0.15 \, , \\
 z & = 1.5 \pm 0.1 \, .
\end{align}
The result is consistent with the value $z=3/2$ 
obtained to the first order in the critical disorder strength in Ref.~\cite{GC}.

\begin{figure}[tb]
  \begin{tabular}{cc}
    \includegraphics[width=40mm]{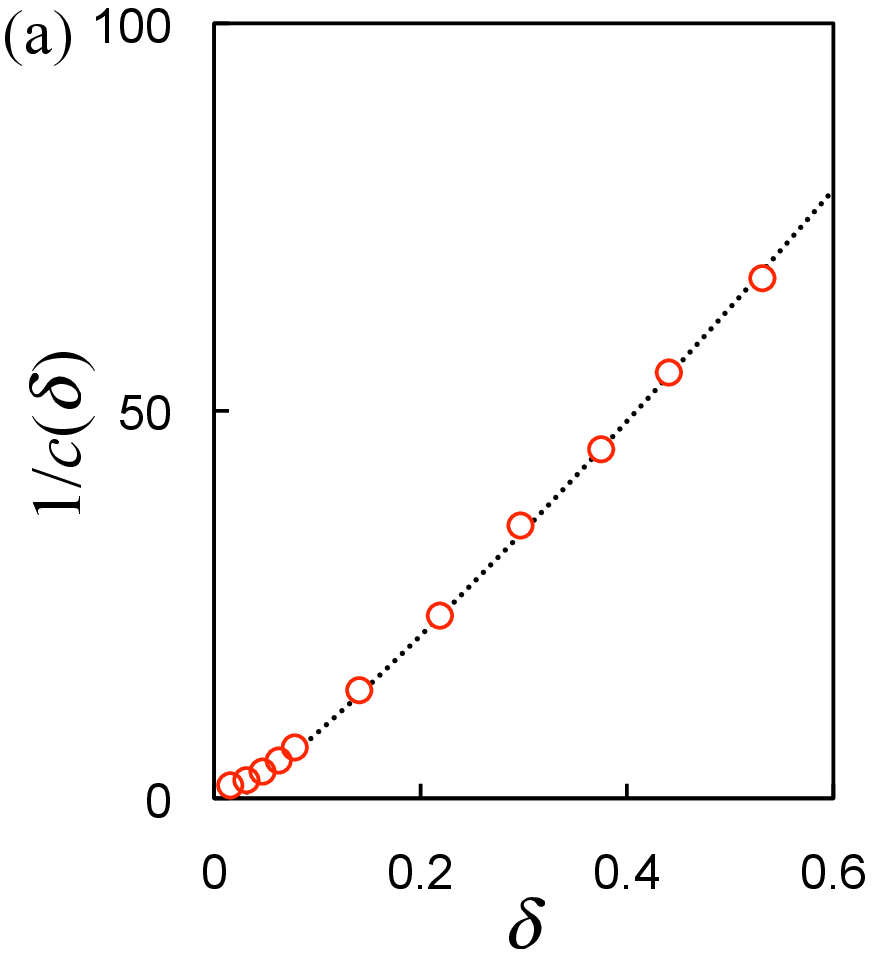}&
    \includegraphics[width=40mm]{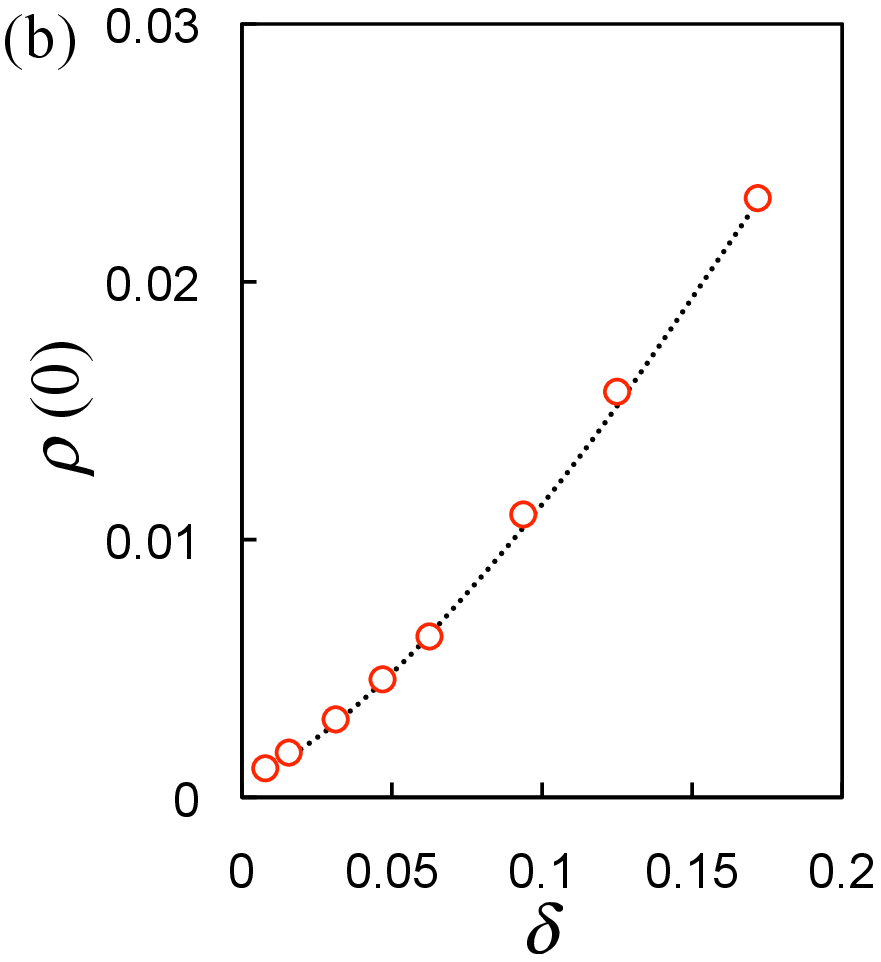}
  \end{tabular}
  \vspace{-2mm}
  \caption{ (Color online) 
     Dependence on $\delta$ 
     (a) for Eq.~\eqn{fit_DSM} on the DSM line 
     and 
     (b) for Eq.~\eqn{DOS_scaling_M} in the ${\bf M}$-phase.
     We set $W_{\rm c}=6.4$.
  }
  \label{fig:fit_DSM_M}
 \end{figure}

 Next we derive the critical exponent $\nu$
from the DOS for small $|\epsilon|$.
 On the DSM line,
by fitting the data to
\begin{align}
 \rho(\epsilon) \sim c(\delta) |\epsilon|^2\, ,
\end{align}
and then by fitting the coefficient $c(\delta)$ to the form
\begin{align} \label{eqn:fit_DSM}
 c(\delta)^{-1} \sim \delta^{3(z-1)\nu_{\rm DSM}}\, ,
\end{align}
we find [Fig.~\ref{fig:fit_DSM_M}(a)]
\begin{align} \label{eqn:nu_DSM}
 3(z-1)\nu_{\rm DSM} &\simeq 1.16 \pm 0.05 \, , \\
 \therefore \  \nu_{\rm DSM} &\simeq 0.81 \pm 0.21 \, .
\end{align}
 The result can be interpreted physically as vanishing velocity of the Dirac electron along the DSM line towards the tricritical point $\delta = 0$,
\begin{align} \label{eqn:velocity}
 v \sim \delta^{(z-1)\nu} \approx \delta^{0.4}\, .
\end{align} 
 In the ${\bf M}$-phase, on the other hand, 
by fitting the data to Eq.~\eqn{DOS_scaling_M},
we find [Fig.~\ref{fig:fit_DSM_M}(b)]
\begin{align} \label{eqn:nu_M}
 (3-z)\nu_{\bf M} &\simeq 1.36 \pm 0.09 \, , \\ 
 \therefore \  \nu_{\bf M} &\simeq 0.92 \pm 0.13 \, .
\end{align}
 The values of $\nu_{\bf M}$ and $\nu_{\rm DSM}$ agree within the margin of error, 
and one expects in fact the same value on both sides of the transition.
 The first order perturbation theory in the location of the critical point \cite{GC} yields the characteristic $\nu_{\rm DSM} = \nu_{\bf M} = 1$, 
which also falls within our intervals on both sides.

\begin{figure}[tb]
  \begin{tabular}{c}
   \includegraphics[width=85mm]{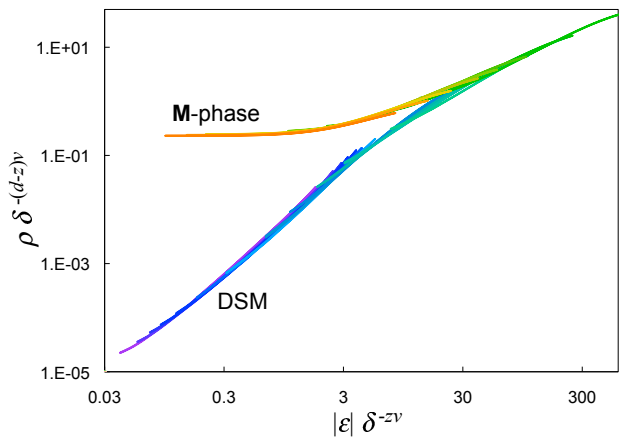}
  \end{tabular}
  \vspace{-2mm}
  \caption{ (Color online)
     Single parameter scaling of the DOS.
     The upper branch corresponds to the DOS in ${\bf M}$-phase, 
    and the lower branch to the DSM line.
    We set the parameters $W_{\rm c}=6.4$, $z=1.5$, and $\nu=0.86=(\nu_{\rm DSM}+\nu_{\bf M})/2$.
  }
  \label{fig:DOS_SPS}
 \end{figure}

Lastly, and most importantly, we show that the single parameter scaling law, Eq.~\eqn{DOS_scaling_delta}, fits successfully all of our numerical data.
 Figure~\ref{fig:DOS_SPS} is the plot of the scaling combination 
$\rho(\epsilon) \delta^{-(d-z)\nu}$ vs. $|\epsilon|\delta^{-z\nu}$,
with the above estimates of $W_\mathrm{c}$, $z$, and with using the average of the two exponents, $\nu=(\nu_{\rm DSM}+\nu_{\bf M})/2 = 0.86$.
 A similar value for $\nu$ would also follow had we solved Eqs.~\eqn{nu_DSM} and \eqn{nu_M} under the assumption that $\nu_{\bf M}=\nu_{\rm DSM}$. 
 After cutting off the relatively large energy region outside the Dirac cone 
and the very small energy region where the DOS becomes too small to estimate numerically,
all the curves in Fig.~\ref{fig:dos} collapse onto two distinct branches,
corresponding to the ${\bf M}$-phase and to the DSM line, respectively.
 This is the central result of the present work.

 The general scaling arguments imply interesting transport properties as well.
 Consider, for example, the wave packet dynamics \cite{tomi1}.
 We assume the mean square displacement $\braket<\bm{r}^2 (t,\epsilon)>$ of the state with energy $\epsilon$
at time $t$, where $\braket<\cdots>$ represents both quantal and ensemble averages to be of the form
\begin{align}
 \braket<\bm{r}^2 (t,\epsilon)> \sim \xi^2 g(t\xi^{-z},|\epsilon|\xi^z)\, .
\end{align} 
 In the ${\bf M}$-phase, one expects $\braket<\bm{r}^2 (t,\epsilon)> = 2dD(\epsilon) t$ for large $t$ with $D(\epsilon)$ the diffusion coefficient at energy $\epsilon$.
 We focus only on the state with $\epsilon=0$,
\begin{align}
 \braket<\bm{r}^2 (t,0)> \sim 
                     \xi^{2-z} t\, ,
\end{align} 
implying the diffusion coefficient $D(0)$ to diverge while the conductivity $\sigma(0) \sim \rho(0) D(0)$ to vanish towards ${\bf P}_{\rm c}$ 
as
\begin{align}
 D(0) \sim \delta^{-(2-z)\nu}\, ,\,\sigma(0)\sim\delta^{(d-2)\nu}\,,
\end{align} 
the latter coinciding with the Wegner's relation \cite{wegner76}, and predicts $\sigma(0)\sim \delta^{0.9}$.
 At ${\bf P}_{\rm c}$, the $\xi$ dependence should vanish, leading to
\begin{align}
 \braket<\bm{r}^2 (t,0)> \sim \xi^2 (t\xi^{-z})^{2/z} = t^{2/z} \approx  t^{1.3} \, ,
\end{align} 
which implies {\it superdiffusion}: 
when $z\simeq 1.5 <2$, the system at ${\bf P}_{\rm c}$ is more diffusive than in the ${\bf M}$-phase.
 The numerical verification of such a superdiffusive behavior is, however, difficult, 
since we need to focus on the wave packet dynamics of $\epsilon=0$ state,
the DOS of which is vanishing.
 Study is in progress to improve the situation.

 Another interesting quantity is the conductance distribution along the DSM line.
 Away from ${\bf P}_{\rm c}$, 
the conductance will be narrowly distributed about the value expected in the absence of randomness 
as demonstrated in Ref.~\cite{KOI}.
 At ${\bf P}_{\rm c}$, we expect the
scale independent broad conductance distribution as in the case of
the Anderson transition \cite{shapiro90,slevin97}.

 In summary,
we have proposed the scaling of the density of states 
as a characteristic of the semimetal to metal transition in general, 
or, of the tricritical point among the two topologically different insulating phases and the metallic phase, in particular.
 In contrast to the conventional Anderson transitions, 
the density of states plays the role of the order parameter and 
shows the universal single-parameter scaling.
 This idea of using DOS to characterize DSM is also relevant in different systems such as the ones reported recently in Refs.~\cite{Nandkishore13} and \cite{Ominato13}.
 Furthermore, we have estimated numerically the dynamical exponent $z \simeq 1.5$, 
which is clearly different from the conventional value $z = 3$ \cite{wegner76} for the Anderson transition in 3D.
 The critical exponent of divergence of the length scale $\nu \simeq 0.9$ is less accurate, 
but it also seems rather far from 
the conventional value $\nu \simeq 1.35$ \cite{asada05} for the Anderson transition in 3D symplectic class.
 The poor inaccuracy of $\nu$ originates from the uncertainty of $W_{\rm c}$ and $z$.
 High precision estimate of $W_{\rm c}$ by different methods such as the transfer matrix \cite{KOI} would improve the estimate.
  
 In this paper, we have focused on the phase boundary of the strong and weak topological insulators.
 The reason is practical; 
the DSM line and the phase boundary of metal to topological insulator phases 
intersect with a large angle, allowing us to pinpoint ${\bf P}_{\rm c}$ easily.
 For the phase boundary of the strong topological and ordinary insulators (STI/OI) \cite{KOI}, 
it is rather challenging to locate ${\bf P}_{\rm c}$, 
because the DSM line and the phase boundary of metal to insulator seem to intersect with a shallow angle.
 Because of the universal nature of critical phenomena,
we expect similar scaling behavior with the same critical exponents for
the semimetal to metal transition for STI/OI.
 On the other hand, different critical behavior is expected for the case of $\mathbb Z$ topological superconductor described by a Bogoliubov-de Gennes Hamiltonian, 
which shows similar phase diagram but belongs to a different universality class (DIII).

\begin{acknowledgments}
 This work was supported by Grants-in-Aid for Scientific Research (C) (Grants No.~23540376)
and Grants-in-Aid 24000013.
 I.~F.~H. is supported by the NSERC of Canada. 
 K.~K. and T.~O. would like to thank Zhejiang Institute of Modern Physics, where fruitful discussion
with V.~E.~Sacksteder and K.~Slevin has been made.
\end{acknowledgments}

\bibliography{KOIH_131210a}

\begin{thebibliography}{39}
\bibitem[{Ord()}]{OrderOfPolynomial}
\bibinfo{note}{We employed a polynomial of the second order for the
  extrapolation. The result is not sensitive to the extrapolation method near
  ${\bf P}_{\rm c}$.}
\end{thebibliography}

\end{document}